\documentclass[12pt]{article}

\usepackage{epsfig}
\usepackage{rotating}

\newcommand{\bcen}{\begin{center}}
\newcommand{\ecen}{\end{center}}

\newcommand{\beq}{\begin{equation}}
\newcommand{\eeq}{\end{equation}}
\newcommand{\beqn}{\begin{eqnarray}}
\newcommand{\eeqn}{\end{eqnarray}}

\newcommand{\bdm}{\begin{displaymath}}
\newcommand{\edm}{\end{displaymath}}

\begin{document}
\begin{center}
{\Large \bf Coupling of Baryon Resonances \\ to the  $N\,\omega$ channel}
\footnote{Work supported by BMBF and DFG.}
\bigskip
\bigskip

{M. Post and U. Mosel}

\bigskip
\bigskip
{ \it
Institut f\"ur Theoretische Physik, Universit\"at Giessen,\\
D-35392 Giessen, Germany \\ and \\
Nuclear Science Division, Lawrence Berkeley 
National Laboratory\\ Berkeley, CA 94720, USA\\
}
\end{center}

\bigskip
\bigskip


\begin{abstract}
We estimate the resonance coupling strength $f_{RN\omega}$ and $f_{RN\rho}$
from a Vector Meson Dominance (VMD)
analysis. The isoscalar and isovector part of the photon coupling are
obtained separately from
helicity amplitudes. The reliability of this approach is tested by comparing
VMD predictions for $f_{RN\rho}$ with values obtained from fitting 
the hadronic decay widths into $N\,\rho$.
A reasonable agreement is found, 
but VMD tends to underestimate the coupling constants. 
In order to confirm consistency with experimental data, we calculate
the cross-sections for photon-and pion induced reactions
within a {\it Breit-Wigner} model. Finally, we study 
how the properties of $\omega$ mesons in nuclear matter 
are affected from the excitation
of resonance-hole loops. For an $\omega$ at rest, 
we find a broadening of about $40$ MeV, while at higher momenta the effect
of  resonance excitations is reduced. \\



\end{abstract}


\newpage


\section{Introduction}

Up to now a satisfactory understanding
of baryon resonances in the $N\,\omega$ channel
has not been achieved. A number of experiments
on the photoproduction of $\omega$ mesons have displayed 
indications for resonant structures 
\cite{burkert}.
However the data basis is not sufficiently broad to allow for an 
unique determination of the quantum numbers of the involved resonances.
Thus -- based mostly on quark models --
a number of resonances in the mass range of 
approximately $1.8-2$ GeV have been proposed
to show a sizeable coupling to the $\omega$ meson \cite{bennhold,bennhold2,capstick}.
On the other hand, no experimental evidence for the existence of 
baryon resonances in the $N\,\omega$ system has been reported from 
the reaction $\pi\,N \rightarrow \omega\,N$  \cite{landolt}.

An understanding of baryon resonances in the $N\,\omega$ channel is not only
important for a description of scattering experiments, but might also be of
interest as far as the properties
of $\omega$ mesons in nuclear matter are concerned.
This conjecture is guided from the experience with the $\rho$ meson, which
has been shown to undergo 
significant modification in the nuclear medium due to the excitation of 
resonance-hole loops. 
In this scenario a central role is played by the 
subthreshold resonance $D_{13}(1520)$ \cite{pp98,plm,frilu},
suggesting that especially
information on the coupling strength of subthreshold resonances to
the $N\,\omega$ system would be very valuable for a determination of the 
in-medium properties of the $\omega$ meson.

In this work we present estimates for the coupling strength 
of baryon resonances to the $\omega\,N$ channel $f_{RN\omega}$.
Vector Meson Dominance (VMD) is utilized to relate $f_{RN\omega}$ to the
isoscalar strength of the electromagnetic decay of the respective resonance,
which is extracted from helicity amplitudes.
The analysis covers all resonances for which
helicity amplitudes are currently available \cite{arndt,pdg}, most of which
are below the nominal $\omega\,N$ threshold.
A coupling of subthreshold resonances to the $N\,\omega$ system
has also been reported elsewhere \cite{frilu,brri}.
In \cite{frilu} such a mechanism is found to be necessary for
a satisfactory description of $\pi\,N$ scattering within
a coupled-channel analysis, whereas 
the authors of \cite{brri} derive the coupling strength within
a quark-model approach. We will compare our results with those of 
both works. 

The paper is organized as follows.
The details of the extraction of $f_{RN\omega}$
are presented in Sect. \ref{coupling}. In Sect. \ref{rhocoupl}
we examine if our model gives reasonable results by comparing the
VMD prediction for the isovector strength
with fits to the hadronic decay width into the $N\,\rho$ channel.
Sect. \ref{omegacoupl} is devoted to a presentation of the results
for $f_{RN\omega}$ and a discussion of their compatibility with experimental 
information on the reactions $\pi\,N\rightarrow \omega\,N$ and
$\gamma\,N\rightarrow \omega\,N$.
Finally, in Sect. \ref{omegamed} the implications of excitations of
resonance-hole loops for the in-medium properties of $\omega$ mesons are 
examined.


\section{Determination of $f_{RN\omega}$}
\label{coupling}

In this section we explain how we obtain an estimate for the magnitude
of the coupling strength of a baryonic resonance to the $N\,\omega$ channel. 
Before going into the details of the calculation, we outline the
basic idea of our approach.

\begin{figure}
\epsfig{file=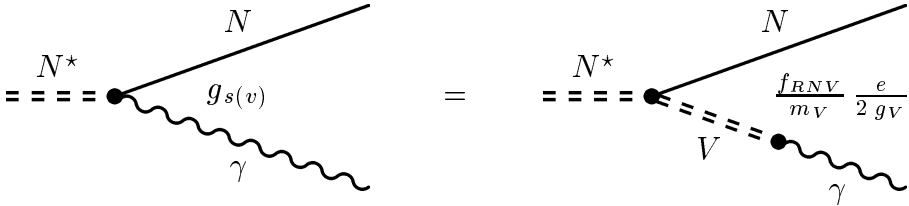,width=12cm}
 \caption{ 
\label{vmd_dec} Diagrammatic description of the electromagnetic decay of 
a resonance in the VMD approach. The diagrams correspond to Eq. \ref{vmdcoupl}.}
\end{figure}

Vector Meson Dominance (VMD) \cite{williams}, a theory which describes
photon-hadron interactions exclusively
in terms of vector meson-hadron interactions,
relates the hadronic coupling strength of resonances to vector mesons 
$f_{RN\rho(\omega)}$ and the isoscalar and isovector part of the photon-coupling,
see Fig. \ref{vmd_dec}:
\beqn
\label{vmdcoupl}
f_{RN\omega} &=& g_s \, m_\omega \, \frac{2\,g_\omega}{e} 
\nonumber \\ && \\ 
f_{RN\rho} &=& g_v \, m_\rho \, \frac{2\,g_\rho}{e} \nonumber \quad. 
\eeqn
As values for $g_\rho$ and $g_\omega$ -- the coupling strengths 
of $\rho$ and $\omega$
meson to the photon -- we take $g_\rho = 2.5$ and  $g_\omega = 8.7$ \cite{williams}. 
The isoscalar and isovector  
coupling strength of the resonance to the $N\,\gamma$ system 
is given by $g_s$ and $g_v$, respectively, see Eq. \ref{isospin}.

Thus VMD gives access to both $f_{RN\omega}$ and $f_{RN\rho}$,
if it is possible to obtain $g_s$ and $g_v$ 
from experimental data.
In order to achieve this goal, it is clearly not sufficient to 
consider merely the total electromagnetic decay width of the resonance. 
Rather, the coupling has to be decomposed into an isoscalar
and an isovector part,
which is readily done by constructing suitable linear combinations
of proton- and neutron-amplitudes. 
These amplitudes are known from experiment \cite{arndt,pdg},
thus allowing to deduce numerical values for $g_s$ and $g_v$.

In a nonrelativistic formulation, the coupling of baryon resonances 
to the photon-nucleon system can be formulated as \cite{pp98,fp}:

\beq
\label{lagrangian}
{\cal L}_{RN\gamma} = \left \{
\begin{array}{lllll}
\chi^\dagger_R\,\epsilon_{ijk}\,\sigma^i\,q^j\,\epsilon^k_\lambda\,\chi_N \,I
+ h.c. & \quad \mbox{for} & J^P = \frac{1}{2}^+ \\ \\
\chi^\dagger_R\,\epsilon_{ijk}\,S^i\,q^j\,\epsilon^k_\lambda\,\chi_N \,I
+ h.c. & \quad \mbox{for} & J^P = \frac{3}{2}^+ \\ \\
\chi^\dagger_R\,R_{ij}\,F^{ij}_\lambda\,\chi_N \,I
+ h.c. & \quad \mbox{for} & J^P = \frac{5}{2}^+  \\ \\
\chi^\dagger_R\,(\sigma_k\,\epsilon^k_\lambda\,\omega - 
\sigma_k\,q^k\,\epsilon^0_\lambda)\,\chi_N \,I
+ h.c. & \quad \mbox{for} & J^P = \frac{1}{2}^- \\ \\
\chi^\dagger_R\,(S_k\,\epsilon^k_\lambda\,\omega - 
S_k\,q^k\,\epsilon^0_\lambda)\,\chi_N \,I
+ h.c.& \quad \mbox{for} & J^P = \frac{3}{2}^- \quad .
\end{array} \right .
\eeq

Here $\chi^\dagger_R$ and $\chi_N$ denote resonance and nucleon spinors
respectively. The $\sigma^i$ are the Pauli matrices, $S^i$
denotes the spin-$\frac{3}{2}$ and $R^{ij}$ the 
spin-$\frac{5}{2}$ transition operator.
$q$ is the c.m. momentum of the photon, $\omega$ its energy and
$\epsilon_\lambda$ its polarization vector. $F^{ij}_\lambda$ denotes the spatial
components of the electromagnetic field strength tensor $F^{\mu\nu}_\lambda$ with
$F^{ij}_\lambda=q^i\,\epsilon^j_\lambda -  q^j\,\epsilon^i_\lambda$. By $I$ we denote
the isospin part of the coupling which is given as:
\beqn
\label{isospin}
I = \chi_R^{I\,\dagger}\, \left( g_s + g_v\,
\left\{ \begin{array}{ll}\sigma^3 \\  S^3 \end{array}\right\}
 \right)\, \chi^I_N  \quad \mbox{for} \quad 
\left\{ \begin{array}{ll} I_R = \frac12 \\ \\  I_R = \frac32 \end{array}\right. \quad .
\eeqn
The spinors $\chi_R^I$ and $\chi_N^I$ represent resonance and
nucleon isospinors and $I_R$ denotes the isopin of the resonance.
$\sigma^3$ and $S^3$ are defined as above. 

Note that the invariant amplitude -- as derived from
(\ref{lagrangian}) and (\ref{isospin}) -- for the decay of a resonance 
into a proton, ${\cal M}_p$,   
is proportional to $g_p = g_s + \alpha\,g_v$, whereas for the neutron the amplitude
${\cal M}_n$  is proportional 
to $g_n = g_s - \alpha\,g_v$. The factor $\alpha$ is defined as:
\beq
\alpha = \chi_R^{I\,\dagger}\,
\left\{ \begin{array}{ll}\sigma^3 \\  S^3 \end{array}\right\}
\, \chi^I_N \quad.
\eeq
For nucleon resonances it is $1$ and for $\Delta$ resonances it is  exactly 
the Clebsch-Gordan coefficient for the respective transition, in this case 
$\alpha = \sqrt \frac{2}{3}$ \cite{eriwei}. 

Thus, the linear combinations
\beqn
\label{scavec}
{\cal M}_{s} &=& \frac{1}{2}\, \left( {\cal M}_p + {\cal M}_n \right) \nonumber \\
             && \\
{\cal M}_{v} &=& \frac{1}{2}\, \left( {\cal M}_p - {\cal M}_n \right) 
\left\{\begin{array}{ccc}
1 &\mbox{for}&I_R = \frac12 \\ \\
\frac{1}{\alpha} &\mbox{for}&I_R = \frac32 \end{array} \right.
\nonumber
\eeqn
are proportional to $g_s$ and $g_v$ respectively.
${\cal M}_{s}$ and ${\cal M}_{v}$ describe the $RN\gamma$
system in terms of the isospin of the photon rather than the isospin of the nucleon.

Experimental information on the decay amplitudes ${\cal M}_p$ and ${\cal M}_n$ exists in
form the of measured helicity amplitudes $A_{\frac12}^{p/n}$ and
$A_{\frac32}^{p/n}$. They 
describe the transition of the photon-proton and the photon-neutron system to a resonance
state with $j_z = \frac{1}{2} \,\mbox{or} \, \frac{3}{2}$, where the
quantization axis is defined along the direction of the photon momentum.
From the helicity amplitudes the isoscalar and isovector part of the coupling are
constructed in exactly the same way as shown in Eq. \ref{scavec}
for the Feynman amplitudes.
The helicity amplitudes are known from experiment at the pole-mass of 
the resonance, allowing the determination of $g_s$ and $g_v$.

To this end we introduce the $\gamma$-width  $\Gamma_{s/v}^{\gamma}$, which -- using the
normalization from \cite{pdg} -- is defined in terms of the helicity amplitudes
$A_{s/v}$ in the following way:
\beq
\label{gammawidth}
\Gamma_{s/v}^{\gamma}(m_R) = \frac{q^2}{\pi}\,
\frac{2 m_N}{(2j_R+1)m_R}\,\left(|A_{\frac12}^{s/v}|^2 + |A_{\frac32}^{s/v}|^2 \right)\quad, 
\eeq
with $j_R$ and $m_R$ denoting spin and pole-mass of the resonance.
Note that after the angular integration has been performed no interference 
terms between $A^{\frac12}$ and $A^{\frac32}$ appears in Eq. \ref{gammawidth}.

Clearly, $\Gamma_{s/v}^{\gamma}$ can also be expressed using Feynman amplitudes:
\beq
\label{gammawidth2}
\Gamma_{s/v}^{\gamma}(k^2) 
= \frac{1}{(2j_R+1)}\frac{q}{8\,\pi\,k^2}\, |{\cal M}_{s/v}|^2 \quad,
\eeq
where $\sqrt{k^2}$ is the invariant mass of the resonance and $q$ the momentum
of the photon in the rest frame of the resonance.
After summing over the photon polarizations,
$|{\cal M}_{s}|^2$ assumes the following form:
\beq
|{\cal M}_s|^2 = 4\,m_N\,m_R\,\kappa \,g_s^2 \,q^2 \, F(k^2) \quad.
\eeq
The same relation holds for $|{\cal M}_v|^2$ with $g_s$ 
replaced by $g_v$. 
For the formfactor $F(k^2)$ at the $RN\gamma$ vertex we assume the following
functional dependence \cite{plm,feupho,feu}:
\beq
\label{formfac}
F(k^2) = \frac{\Lambda^4}{\Lambda^4 + (k^2-m_R^2)^2} 
\eeq
with $\Lambda=1$. If we consider on-resonance decays ($\sqrt{k^2}=m_R$) this formfactor
is equal to $1$. The numerical factor $\kappa$ depends on the
quantum numbers of the resonance. For isospin-$\frac{1}{2}$ resonances the values 
for $\kappa$ are listed in Table
\ref{vmdom}. In the case of $\Delta$ resonances they need to be multiplied
by a factor of $F^2 = \frac{2}{3}$ due to isospin.  
The two expressions Eqs. \ref{gammawidth} and \ref{gammawidth2}
can now be equated allowing to solve for $g_{s/v}$:
\beq
\label{resg}
g_{s/v}^2 = \frac{4}{\kappa} \, \frac{|A_{\frac12}^{s/v}|^2 +
|A_{\frac32}^{s/v}|^2}{q} 
\left\{\begin{array}{ccc}
1 &\mbox{for}&I_R = \frac12 \\ \\
\frac{1}{F^2} &\mbox{for}&I_R = \frac32 \end{array} \right. \quad.
\eeq
Again, in the case of $\Delta$ resonances an additional factor $\alpha^2$ needs
to be introduced. 

Thus it is possible to obtain $g_{s/v}$ from helicity amplitudes. 
The hadronic couplings $f_{RN\omega(\rho)}$ 
are then readily deduced from the VMD relation Eq. \ref{vmdcoupl}. The corresponding
values are listed in Tables \ref{vmdrho} and \ref{vmdom}.

A modified version of VMD 
(called VMD1 in \cite{williams}) contains also a direct coupling
between hadrons -- in our case baryon resonances -- and the photon. 
In particular, at the photon point $q^2 = 0$ the photon-resonance
interaction proceeds without intermediate vector mesons, 
whereas the decay into a massive photon consists
of both a direct coupling term and a term with an intermediate vector meson. 
In principle, it would be preferable to perform an analysis such as ours 
within the framework of the modified version of VMD by studying 
-- currently unavailable -- dilepton production data on the nucleon. 
This process involves massive photons and would, in combination with
the data for real photons, allow one to disentangle the direct photon and vector meson
contribution. As a further advantage, the
coupling strength of the vector mesons then does not need to be extrapolated 
over the large mass range 
from the photon point to the on-shell mass of $\rho$ and $\omega$ meson.


\section{Results for the $\rho$: How Reliable is VMD ?}
\label{rhocoupl}

In this section we study the applicability of VMD in the resonance region.
This is done for the isovector part of the analysis, where the
VMD predictions for $f_{RN\rho}$ can be compared with the measured 
resonance decay width
into the $N\,\rho$ channel.

We consider all resonances with $j_R < \frac72$, 
for which both the hadronic decay widths into $N\,\rho$ and the helicity
amplitudes are known. This excludes a few light resonances like the $S_{11}(1535)$
and the $P_{33}(1600)$, for which only upper limits for the $N\,\rho$
branching ratio exist \cite{pdg}. 
For both the helicity amplitudes and for the partial $N\,\rho$ decay width
we use different parameter sets in order to
provide an estimate for the experimental uncertainties entering this analysis.
The helicity amplitudes are taken from Arndt {\it et al} \cite{arndt} and the PDG
\cite{pdg}. 
Furthermore we use the values obtained in the work of \cite{feupho}, 
which is of particular interest as it represents
the only available approach which describes simultaneously hadronic-and 
electromagnetic reactions over the full energy region of interest here. 
The $\rho$ decay widths are taken from 
the analysis of Manley {\it et al} \cite{man1,man2} and the PDG \cite{pdg}. 
Furthermore, we study the 
additional model dependence introduced by different parameterizations of the 
spectral function of the $\rho$ meson. 

The hadronic coupling constant $f_{RN\rho}$
is obtained via the following expression for the $R \rightarrow N\,\rho$ 
decay width (taken on the pole mass of the resonance)\cite{pp98}:
\beq
\label{rhowidth}
\Gamma_{R\rightarrow N\,\rho} = \frac{1}{8 \pi m_R^2} \, 
\frac{1}{2 j_R + 1} \!\!\!\!
\int\limits_{2 m_\pi}^{m_R - m_N} \!\!\!\! dm \, 2m \, A_\rho(m)\, 
|{\cal M}_{R\rightarrow N\,\rho}|^2 \, {\bf q} \quad.
\eeq

The decay amplitude ${\cal M}_{R\rightarrow N\,\rho}$
is obtained from the same Lagrangians as in the electromagnetic case
and is proportional to $f_{RN\rho}$ \cite{pp98}.
${\bf q}$ is the three momentum of the $\rho$ meson in the rest frame 
of the resonance. Note that there is no formfactor in Eq. \ref{rhowidth}
because we consider the on-resonance decay.

The spectral function of the $\rho$ meson $A_\rho(m)$ is taken as:
\beq
A_\rho(m) = \frac{1}{\pi} \, \frac{m\,\Gamma(m)}{(m^2-m_\rho^2)^2 +
m^2\,\Gamma^2(m)} \quad .
\eeq
The decay into $N\,\rho$ of low-lying resonances, for example the $D_{13}(1520)$, 
proceeds mainly through the low-mass tail of $A_\rho$.
The shape of the tail is quite sensitive to the parameterization
of the $\rho$ decay width. To study the effect on $f_{RN\rho}$ we
compare two different parameterizations of the $\rho$ decay width:
\beqn
\label{rhowidth1}
\Gamma(m) &=& \left(\frac{m_\rho}{m}\right)^2\,\Gamma_0\,
\left(\frac{{\bf p}_m}{{\bf p}_{m_\rho}} \right)^3  \\
\mbox{and} && \nonumber \\
\label{rhowidth2}
\Gamma(m) &=& \left(\frac{m_\rho}{m}\right)\,\Gamma_0\,
\left(\frac{{\bf p}_m}{{\bf p}_{m_\rho}} \right)^3 \,
\frac{1 + r^2\,{\bf p}_{m_\rho}^2}{1 + r^2\,{\bf p}_{m}^2} \quad .
\eeqn
The first version follows from a one-loop approximation of the self energy of the 
$\rho$ meson \cite{pp98,klingl}, the second one is taken from the PDG \cite{pdg}.
$m$ is the invariant mass of the $\rho$ meson, $m_\rho = 0.768$ GeV its pole mass and
$\Gamma_0 = 150$ MeV its decay width
on the pole mass. ${\bf p}_{m_\rho}\,({\bf p}_m)$ 
denotes the 3-momentum of the pions measured
in the rest frame of a decaying $\rho$ meson with mass $m_\rho\,(m)$.
The range parameter $r$, appearing in the second expression has the numerical
value $r=5.3$/GeV \cite{pdg}. 

In Table \ref{vmdrho} the results for the coupling constants and the
corresponding error-bars - which are calculated from the error-bars assigned to
the partial decay width and the helicity amplitude in the respective analysis
- are given. As a general tendency, our finding is that VMD
works well within a factor of two. 
It tends to somewhat underestimate the
coupling constants, leaving some room for an additional direct coupling of the
resonance to the photon. This can be seen particularly well in the case of the
$D_{13}(1520)$ and the $F_{15}(1680)$, which are the most prominent resonances
in photon-nucleon reactions, and whose $\rho$ decay widths are also well
under control. Note that in the case of the $S_{31}(1620)$ and the $P_{13}(1720)$ 
large discrepancies occur between the VMD predictions
based on different sets of helicity amplitudes, 
thus reflecting the uncertainties inherent in this analysis. The helicity 
amplitudes from Arndt \cite{arndt} and PDG \cite{pdg} produce nearly 
identical results for the $\rho$ coupling of the $P_{13}(1720)$, which might 
seem suprising as the helicity amplitudes 
from both works are very different. However, most of these differences cancel out
in the isovector channel
after proton- and neutron- amplitudes have been subtracted from each other (see Eq. \ref{scavec}).
In the isoscalar sector this cancellation does not occur and 
both sets lead to very different predictions for $f_{RN\omega}$, see Sect. 
\ref{omegacoupl}.

For the $P_{13}(1720)$ and the $F_{35}(1905)$ VMD is off by an order 
of magnitude. We argue that this mismatch does not 
hint on a failure of VMD, but has to be attributed to the unsatisfactory
experimental information on these two resonances. Comparing the helicity amplitudes
of the $P_{13}(1720)$ obtained from different analyses 
\cite{pdg,feupho} reveals that neither their sign nor their magnitude are
under control at all. 
Also the partial decay width into the $\rho\,N$ channel
is subject to large uncertainties, here the PDG and Manley differ by about 
a factor of three \cite{pdg,man1,man2}.
Obviously, the experimental observation of this resonance
is very complicated and its parameters might be sensitive  
to the details of the underlying theoretical model, such as the treatment
of the non-resonant background. For a conclusive VMD analysis of these
resonances it is therefore mandatory to enlarge the data base and to 
describe hadronic- and photoinduced reactions within one model.
In \cite{fp} a VMD analysis for exactly these resonances was performed, 
leading to the general conclusion - in contradiction to this work - 
that VMD is not at all reliable in the resonance region.

Different parameterizations of the $\rho$ decay width mainly influence
the coupling constants of low-lying resonances.
Since the PDG parameterization of the $\rho$ decay (see Eq. \ref{rhowidth2})
distributes more strength at invariant mass below the $\rho$ meson pole mass,
it leads to smaller values for the coupling constant, thus leading to 
an improved agreement between VMD and the hadronic fits.
This effect is most notable for the $D_{13}(1520)$ (see fourth and
fifth column in Table \ref{vmdrho}).

Summarizing the results, we conclude that VMD can be applied in the
resonance region on a phenomenological basis. 
We have shown that for the coupling constants $f_{RN\rho}$
it leads to an approximate agreement with values adjusted to the 
hadronic decay width $\Gamma_{RN\rho}$ with a tendency to 
somewhat underestimate these values.  
Therefore, our approach should yield reasonable predictions -- at least for the 
lower limits -- of the unknown coupling constants $f_{RN\omega}$.


\section{Results for the $\omega$: Strong Coupling of Subthreshold Resonances }
\label{omegacoupl}

In this section we present our results for the coupling constants
$f_{RN\omega}$ following from the VMD analysis and discuss 
their compatibility with experimental information
obtained from pion- and photon-induced $\omega$-production cross sections.

All nucleon resonances with $j_R < \frac72$, for which helicity amplitudes
have been extracted, are included. Thus we consider only one resonance above
the $N\,\omega$ threshold in our analysis, namely the $D_{13}(2080)$.
Three different parameter sets for the helicity amplitudes are used,
the results from the analysis of Arndt {\it et al} \cite{arndt} and Feuster {\it et al}
\cite{feupho}  as well as the values listed
in the PDG \cite{pdg}. The corresponding results for $f_{RN\omega}$ 
together with the error-bars are given in 
Table \ref{vmdom}. We find a strong coupling to the
$N\,\omega$ channel in the $S_{11}$, $D_{13}$ and $F_{15}$
partial waves; especially the $S_{11}(1650)$, the $D_{13}(1520)$ and the
$F_{15}(1680)$ resonances show a sizeable coupling strength to this
channel. In most cases the three parameter sets for the helicity amplitudes
lead to similar results. As in the case of the $\rho$ meson, however, 
for a few resonaces ($P_{11}(1440)$, $S_{11}(1650)$ and $P_{13}(1720)$)
the differences are more pronounced.
In particular, for the $P_{13}(1720)$ the poor experimental situation does not allow
a reliable extraction of the coupling strength, thus emphasizing the
difficulties concerning this resonance which were already discussed in
the previous section.
The relatively large coupling constant for the $D_{13}(1520)$ might
at first seem surprising as the helicity amplitudes suggest that its
electromagnetic decay is mainly an isovector one,
corresponding to a small value for $g_s$. However,
$f_{RN\omega}$ is proportional to $g_\omega$ (see Eq. \ref{vmdcoupl}),
which is about three times larger than $g_\rho$. As a result, the $\rho$ and
$\omega$ coupling are of the same size, but with a much larger error-bar for the
$\omega$ coupling.

It is noteworthy that the resonances with the largest coupling 
are well below the $N\,\omega$ threshold.
Subthreshold resonances in the $N\,\omega$ channel 
are also reported elsewhere \cite{frilu,brri}. 
In \cite{frilu} it is shown that within a coupled-channel analysis 
the description of $\pi\,N$ scattering enforces 
resonant structures in the $N\,\omega$ channel. As in our work, 
strong contributions from the $S_{11}$ and $D_{13}$ partial waves
are found. Quantitatively, both approaches yield quite different
predictions for the coupling strength of the $D_{13}(1520)$, however.
Whereas our VMD analysis predicts a coupling strength of about $3$ (see Table 
\ref{vmdom}), in \cite{frilu} a value of $f_{RN\omega} \approx 6.5$ is given.
At the same time it is found in 
a quark model calculation \cite{brri} that 
the coupling constant of this resonance should be $f_{RN\omega} \approx 2.6$,
which is surprisingly close to our result. However, from their quark model calculation
the authors of \cite{brri} obtain initially coupling constants for hadronic
Lagrangians which display very different energy dependences compared to ours. 
To be specific, in the case of the $D_{13}(1520)$
close to threshold $\omega$ and nucleon
couple in a relative $d$-wave in their formalism, rather than forming 
an $s$-wave state as following from our Lagrangian, 
see Eq. \ref{lagrangian} (which for example has also been employed in \cite{frilu}). 
Thus a direct comparison of the results as done in \cite{brri} is possible 
only after simplifying assumptions, if at all.

A direct experimental
observation of a resonant coupling of subthreshold resonances is 
very complex, since the resonances can
add to the cross-sections only through the high-mass tails of their mass
distribution and are therefore hard to disentangle from background effects.
Sensitivity to the contribution of subthreshold resonances 
in $\omega$ production experiments can probably
only be achieved by studying polarization observables. The current data 
do not permit such a project, however.
As was pointed out in various previous works \cite{brat,brat2,frisoy2}, 
an analysis of dilepton 
production on the nucleon in combination with VMD
might provide further insight on this issue.

In spite of these difficulties, it is still rewarding to compare
the VMD predictions with existing data on the reactions
$\pi^-\,p \rightarrow \omega\,n$ and $\gamma\,p \rightarrow \omega\,p$.
Thus it is possible to infer if the predicted coupling constants
of the subthreshold resonances produce a satisfactory overall agreement
with data above the threshold. 
Comparison with data allows also to 
discuss the results for the $D_{13}(2080)$, the only resonance in our
analysis above threshold. 
Based on recent data from photoproduction experiments, which display a
richer structure than expected within a simple meson-exchange model,
the search for resonances in the $N\,\omega$ channel has so far concentrated
on the mass range of approximately $1.8-2$ GeV.
At the current stage the data do not allow
to pin down the quantum numbers of the involved resonances, however, and
a variety of predictions exist 
\cite{burkert,bennhold,bennhold2,capstick}. The $D_{13}(2080)$
is not amongst them.
In contrast to the photoproduction data, 
no experimental signature of a coupling of  baryon 
resonances to the $N\,\omega$ channel above the threshold
has been detected in pion induced reactions \cite{landolt}.
We find for the $D_{13}(2080)$
an $\omega$ decay width of about $70$ MeV  and argue 
that its contribution to both reactions is too small to be seen in experiment. 

As a first approximation, we take the full production amplitude
as an incoherent sum of {\it Breit-Wigner} type amplitudes,
describing $s$-channel contributions:
\beq
\label{omsigtot}
\sigma_{ab} = \sum_R \frac{2j_R+1}{(2j_N+1) \Omega_a} \, \frac{\pi}{k_a^2}
\frac{\Gamma_a \, \Gamma_b}{(E_a-E_R)^2 + \Gamma_{tot}^2/4} \quad . 
\eeq

Here $a$ stands for the incoming channel ($\pi^-\,p$ or $\gamma\,p$) and $b$ for
the $\omega\,N$ channel. $\Gamma_{a(b)}$ denote the respective partial decay width of
the resonance and $\Gamma_{tot}$ its total decay width, taking into account the
$\omega\,N$ decay as well. 
The energy dependence
of the photon and the $\omega$ decay amplitudes is obtained from the Lagrangians given
in Eq. \ref{lagrangian} and the formfactor of Eq. \ref{formfac} with  $\Lambda = 1.0$. 
Whenever possible, $f_{RN\omega}$ is obtained from the
helicity amplitudes from Arndt {\it et al}, otherwise the PDG
estimates are used. For the remaining channels we take the
parameterization from Manley {\it et al}. In Eq. \ref{omsigtot} $\Omega_a$ is $1$ for
the pion and $2$ for the photon. $k_a$ is the cm-momentum of the particles in
channel $a$ and $E_a$ denotes their cm-energy. All resonances listed
in Table \ref{vmdom} are included. We stress that the only free
parameter is the cutoff parameter $\Lambda$. 

\begin{figure} 
\epsfig{file=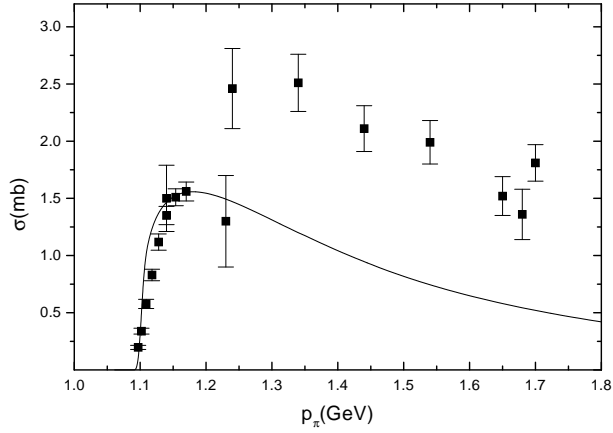,width=10cm}
 \caption{ 
\label{pitot} Total cross-section for the reaction $\pi^-\,p \rightarrow \omega\,n$.
The data are taken from \cite{landolt}.}
\end{figure}

\begin{figure} 
\epsfig{file=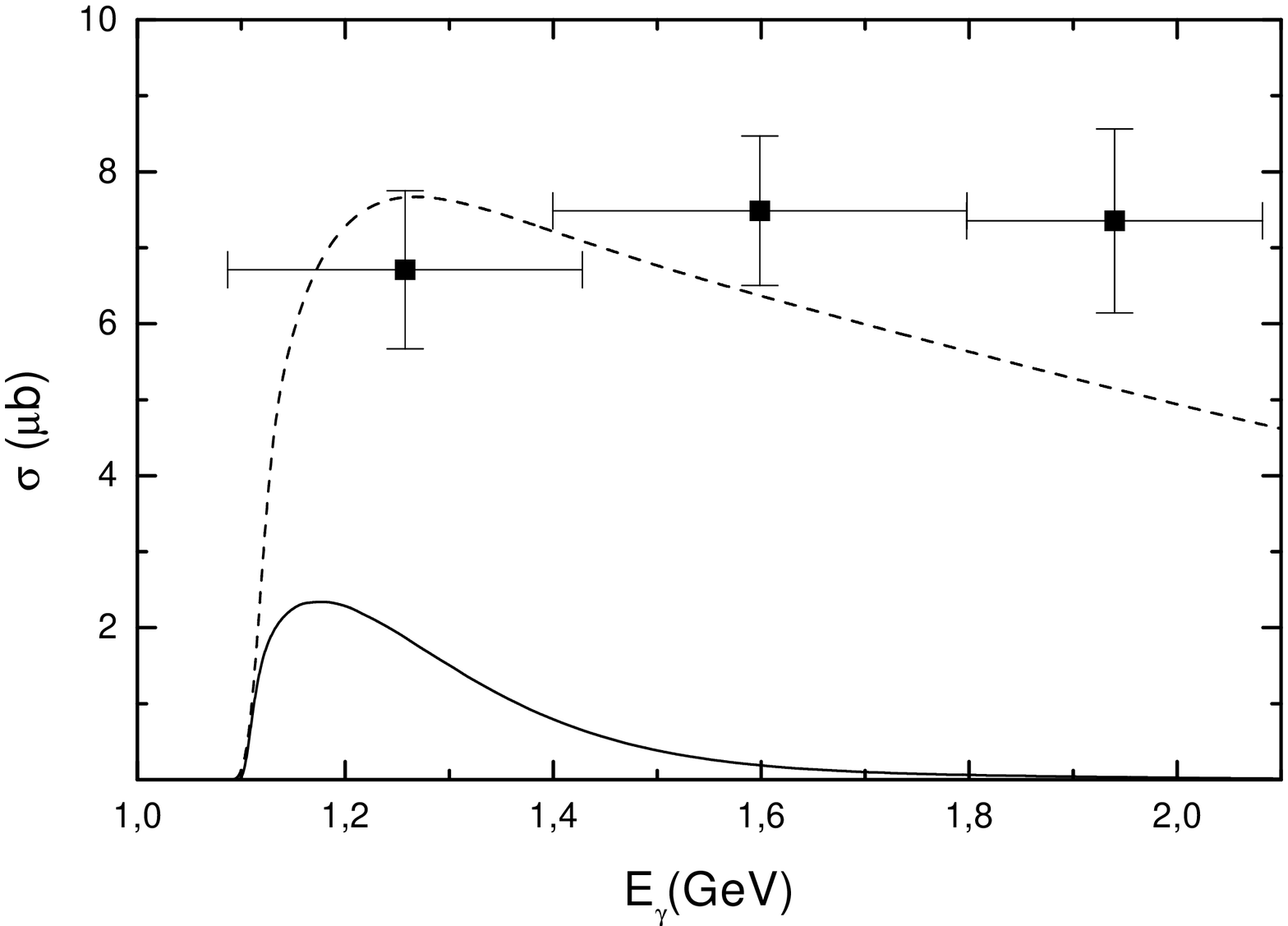,width=10cm}
 \caption{ 
\label{gamtot}Total cross-section for the reaction $\gamma\,p \rightarrow \omega\,p$.
The data are taken from \cite{abbhhm}.The straight line shows the cross-section
resulting from the resonance model, where the dashed line indicates the sum of
resonance and pion-exchange contributions.}
\end{figure}

The results for both reactions are shown in Fig. \ref{pitot} and Fig.
\ref{gamtot}. The data for the $\pi$-induced reaction are taken 
from \cite{landolt} and we use the photoproduction data of \cite{abbhhm}.
The cross-section for $\pi^-\,p \rightarrow \omega\,n$ is 
reproduced rather well, especially near threshold, 
whereas at high energies some strength is missing.
This is in approximate agreement with the findings of the authors of \cite{frilu}, who
are able to give a satisfactory description of this process around
threshold in terms of subthreshold resonances. 
One can therefore conclude that the excitation of subthreshold resonances
constitutes an essential ingredient to the production mechanism. This
interpretation is confirmed by the fact that it is hard to understand
this process in terms of non-resonant amplitudes only.
Already the contribution from $\rho_0$ exchange overestimates the data;
only by suppressing the amplitudes by the introduction of very restrictive 
form factors a rough
agreement with experimental data can be achieved \cite{sibi}. Similar problems have
been reported in \cite{klingl} in a model that includes all Born terms.
The contribution coming from the only resonance above threshold -- the 
$D_{13}(2080)$ -- is about $0.1$ mb, roughly
$10\%$ of the total cross-section. This is certainly too small to 
produce a distinguishable resonant structure in the total cross-section.

On the other hand, the photoproduction data can not be saturated within
the resonance model. In particular, it seems futile to look
for the $D_{13}(2080)$ in this reaction.
Since it is commonly assumed that the $\pi^0$-exchange as invoked in
\cite{frisoy} plays a key role in the photoproduction we -- incoherently --
added this contribution using the same parameters as given in \cite{frisoy}. 
As shown in Fig. \ref{gamtot} the sum of both mechanisms yields a qualitative
explaination of the data over the energy range under consideration.

It is not surprising that the resonance contribution is more likely
to produce lower bounds for the cross-sections. We already discussed 
in Sect. \ref{rhocoupl}, that the VMD analysis tends to underestimate
the hadronic coupling constants. Also, at energies above the $\omega\,N$ threshold,
the helicity amplitudes of only a few resonances are known and the 
experimentally observed cross-sections have to be explained by additional
production mechanisms, such as the $\pi$-exchange in the photoproduction.
Keeping this in mind, the predictions of the resonance model
are in reasonable agreement with the data and can be viewed as a confirmation
of the VMD analysis.


\section{The $\omega$ Meson in Nuclear Matter}
\label{omegamed}

In the previous section we discussed the relevance 
of resonances in the $\omega\,N$ channel
for an understanding of $\omega$ production data. Here we
study up to which extent the existence 
of these resonant states affects the properties of 
$\omega$ mesons in nuclear matter. 
Of particular interest in this context is the fact that the VMD
analysis predicts a strong coupling of subthreshold resonances,
since in the case of the $\rho$ meson it is widely acknowledged that
its in-medium properties are dominated by low lying resonances, especially
the $D_{13}(1520)$.

The in-medium properties of the $\omega$ meson can be read off its 
spectral function, which is defined as:
\beq
A_\omega^{med}(\omega,{\bf q}) = \frac1\pi \, \frac{\mbox{Im}\, \Sigma_{tot}(\omega,{\bf q})}
{(\omega^2-{\bf q}^2-m_\omega^2-\mbox{Re}\, \Sigma_{med}(\omega,{\bf q}))^2 
+ \mbox{Im}\, \Sigma_{tot}(\omega,{\bf q})^2} .
\eeq
Energy and three-momentum of the $\omega$ meson are denoted by 
$\omega$ and ${\bf q}$.
The total self energy of the $\omega$ meson $\Sigma_{tot}$ is the 
sum of the vacuum and the in-medium self energies $\Sigma_{vac}$ and $\Sigma_{med}$. 
In this work we neglect $\mbox{Re}\,\Sigma_{vac}$ and approximate
$\mbox{Im}\,\Sigma_{vac}$ with the dominating $3\,\pi$ decay width. 
We assume that this decay may only proceed via an intermediate $\rho$ meson.
The coupling constant $f_{\omega\rho\pi}$ is adjusted to reproduce the vacuum
decay width of $8.41$ MeV \cite{pdg}.

To lowest order in the nuclear density $\rho_N$, the in-medium self energy of
the $\omega$ meson $\Sigma_{med}$ in symmetric nuclear matter is given by:
\beq
\Sigma_{med} = \rho_N \, T_{\omega\,N} ,
\eeq
where $T_{\omega\,N}$ is the spin/isospin averaged
$\omega\,N$ forward scattering amplitude.
We approximate the scattering amplitude
as a sum over all resonances which are discussed in this work
and determine the resonance contributions in a non-relativistic approach.

For further details of the calculation the reader may consult our previous
publication on the $\rho$ meson \cite{pp98,plm}. 
Both calculations are carried out within the same framework, differences
appearing only in some minor points:
First, the explicit values for the coupling constants and isospin factors
are different. More crucial, we now evaluate the self energy
in the cm-frame rather than in the rest frame of the nucleon.
We do so since it has been demonstrated in \cite{plm} that thus
a much better approximation of a relativistic calculation is achieved.
As explained in \cite{plm}, this leads 
in general to a reduction of the results.
Finally, at the $\omega\,N\,R$ vertex  the same formfactor is taken 
as in Eq. \ref{formfac}, in contrast to the choice in \cite{pp98}.

\begin{figure} 
\epsfig{file=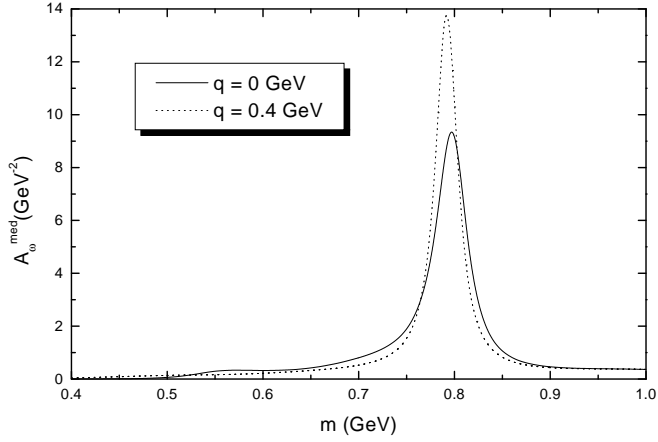,width=10cm}
 \caption{ 
\label{spec0a4} The spectral function of an $\omega$ meson in nuclear matter 
versus its the invariant mass $m$.
Shown are results at $q=0$ GeV (solid line) and at $q=0.4$ GeV (dotted line).
Only the transverse part is displayed. As discussed in the text, there is
virtually no difference between transverse and longitudinal spectral function
at $0.4$ GeV. }
\end{figure}

\begin{figure} 
\epsfig{file=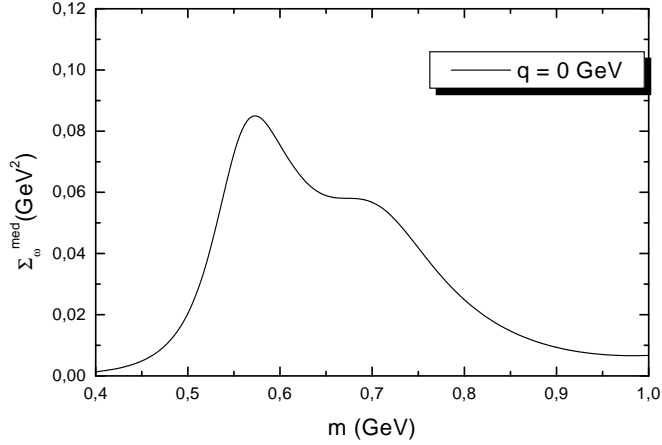,width=10cm}
 \caption{ 
\label{sig0} In-medium self energy of an $\omega$ meson at rest as a function
of its invariant mass $m$. One clearly
sees resonant structures from the excitation of the $D_{13}(1520)$ and
the $S_{11}(1650)$.}
\end{figure}

In Fig. \ref{spec0a4} we show the spectral function for both an $\omega$ 
meson at rest and moving with a momentum of $0.4$ GeV with respect to
the nuclear medium. The most obvious observation
is that our model predicts an in-medium $\omega$ meson which survives very
well as a quasi-particle. 
At rest, its peak position is slightly shifted upwards by
about $20$ MeV, due to level-repulsion.
The in-medium width 
is roughly $40$ MeV as can be read off Fig. \ref{sig0}, where $\Sigma_{med}$
is depicted for an $\omega$ at rest. The $\omega\,N$ forward scattering
amplitude contains all elastic and inelastic channels. In the previous section
we demonstrated that our model can give a good description of 
the important $\pi\,N$ channel close to threshold. This enhances
our belief that the VMD approach yields reasonable results for the 
in-medium properties of the $\omega$ meson, especially for an $\omega$ at rest.

By construction, the 
in-medium self energy develops resonant structures in the
subthreshold region. The most important contributions come 
from the $D_{13}(1520)$ and the $S_{11}(1650)$ resonances.
In contrast to the case of the $\rho$ meson, these structures translate 
only into small bumps in the spectral function, reflecting 
- as follows from the coupling constants - 
that the $\omega$ self energy is  substantially smaller
than that of the $\rho$ meson.

Our value for the in-medium broadening of the $\omega$ meson is 
in approximate agreement 
with that obtained in \cite{klingl} and -- in a dynamical simulation --
in \cite{weid} and \cite{effe}.
Also in the coupled channel analysis of \cite{frilu} an in-medium broadening of $40$ MeV 
for the $\omega$ meson is reported.
However, in this work the authors find that the $\omega$ 
spectrum displays a richer structure in the subthreshold region, 
which is not surprising since their
coupling constant for the $D_{13}(1520)$ is much larger.

At higher momenta the influence of resonance-hole
excitations gets reduced, see Fig. \ref{spec0a4}. Also, 
the longitudinal and transverse channel display a very similar
behaviour (we therefore only show the results for the transverse channel).
This is different from the case
of the $\rho$ meson, whose spectral function 
at large momenta receives a strong broadening 
in the transverse channel from the the coupling of 
$p$-wave resonances, for example the $P_{13}(1720)$,
whilst the longitudinal mode is only
slightly modified. 
In the case of the $\omega$ meson however, only the $F_{15}(1680)$ 
shows a sizeable coupling strength.

We have already mentioned that the $\omega$ decay in vacuum proceeds mainly
via an intermediate $\rho$ meson. Due to kinematics, only the 
low-mass tail of the $\rho$ mass spectrum is involved in this process.
Consequently, if a lot of $\rho$ strength is shifted to 
smaller invariant masses in the nuclear medium -- as is generally believed, 
see for example \cite{pp98,plm,urban} --
the $\omega$ meson will receive additional broadening. 
We have estimated this effect
by replacing the vacuum spectral function of the $\rho$ 
with the in-medium one as given in \cite{plm}  
for the calculation of the $\omega$ decay. 
To lowest order in the density, this corresponds to
scattering processes like $\omega\,N \rightarrow \pi\,R$, where $R$ stands
for any baryon resonance included in the calculation in \cite{plm}; such
processes have so far not been included in any calculation.
We find an additional in-medium width for the $\omega$ 
of about $40$ MeV at the $\omega$ mass from this process. 
It has to be pointed out, however, that
we did not take into account the corresponding mass-shift of the $\omega$ meson.
Since the $\omega$ decay width varies rapidly as a function of its mass, this is a
strong simplification.


\section{Summary}

We have calculated the coupling strength of baryon resonances to vector mesons
in a VMD approach. The isoscalar and isovector strength of the photon coupling 
are extracted by means of helicity amplitudes. For the isovector part
we compare the VMD approach with fits to the  
hadronic decay width of the resonance into the $N\,\rho$ channel.
This allows to test the reliability of VMD in the resonance region. We find
that VMD works well within a factor of two, with a tendency to 
underestimate the coupling strength.

In the $\omega$ channel we find a strong coupling of subthreshold resonances,
especially the $D_{13}(1520)$ and the $S_{11}(1650)$. A direct
experimental observation
of these couplings is very complex and not possible within the current data base,
which provides only an upper bound for the resonance contribution.
We have calculated the total cross-section
for the reactions $\gamma\,N\rightarrow \omega\,N$
and $\pi\,N\rightarrow \omega\,N$ 
with a simple {\it Breit-Wigner} ansatz and have shown that   
our coupling constants do not overestimate the data. In particular, the pion-induced
reaction is very nicely described near threshold. We have argued that
these results can be interpreted as a confirmation of the VMD model.

Guided by the observation, that the $\rho$ meson properties in nuclear matter
are strongly modified from the excitation of resonance-hole loops, we 
estimate these effects for the $\omega$ meson. Since on average the $\omega$
couples weaker to baryon resonances than the $\rho$, the effects are 
not as pronounced. In contrast to the $\rho$ meson,
the $\omega$ survives very well as a quasi-particle.
For an $\omega$ at rest, we find an upward mass-shift of about  
$20$ MeV and an in-medium broadening of about $40$ MeV. At higher momenta
the medium modifications are further reduced.
By taking into account also the possibility that in nuclear matter the $\omega$
meson might decay into a modified $\rho$ meson, we calculate an additional
broadening of the $\omega$ of about $40$ MeV.

\section{Acknowledgements}

The authors gratefully acknowledge many stimulating discussions with V. Koch 
and the hospitality of the Nuclear Science Division of the
LBNL, Berkeley, where parts of this work were completed.
This stay was made possible through support by the NATO science program,
NATO-collaborative research grant number 970102.


\begin{sidewaystable}
\centering
\begin{tabular}{|l||c|c|c||c|c|c|} \hline
& $f_{RN\rho}$(Arndt) & $f_{RN\rho}$(Feuster) & $f_{RN\rho}$(PDG) &
$f_{RN\rho}(Man1)$ & $f_{RN\rho}$(Man2) & $f_{RN\rho}$(PDG)\\ \hline
$D_{13}(1520)$ & 3.44$\pm$0.18 &  2.67 & 3.42$\pm$0.20 
& 6.67$\pm$0.78 & 5.79$\pm$0.68 & 6.66$\pm$1.26 \\ \hline
$S_{31}(1620)$ & 0.89$\pm$0.42 &  0.10 & 0.69$\pm$0.23 
& 2.14$\pm$0.30  & 2.06$\pm$0.28 & 2.61$\pm$1.01 \\ \hline
$S_{11}(1650)$ & 0.70$\pm$0.08 &  0.59 & 0.57$\pm$0.31 
& 0.47$\pm$0.19 & 0.45$\pm$0.18 & 0.79$\pm$0.31 \\ \hline
$F_{15}(1680)$ & 3.48$\pm$0.39 &  ---  & 3.14$\pm$0.37 
& 6.87$\pm$1.57 & 6.36$\pm$1.45 & 7.33$\pm$2.29 \\ \hline
$D_{33}(1700)$ & 3.96$\pm$0.77 &  3.68  & 4.02$\pm$0.62 
& 1.96$\pm$0.67 & 1.93$\pm$0.67 & 4.91$\pm$1.57 \\ \hline
$P_{13}(1720)$ & 0.25$\pm$0.42 &  0.93 & 0.19$\pm$0.79 
&13.17$\pm$3.35 & 12.09$\pm$3.17 & 7.42$\pm$1.64 \\ \hline
$F_{35}(1905)$ & 2.47$\pm$0.55 &  --- & 2.56$\pm$0.75 
& 17.97$\pm$1.14 & 17.57$\pm$1.12 & 14.19$\pm$3.10 \\ \hline
$P_{33}(1232)$ & 13.40$\pm$0.2 & 11.96 & 13.29$\pm$0.28 
& --- & --- & ---- \\ \hline
\end{tabular}
\caption{Listed are the hadronic coupling constants $f_{RN\rho}$ 
obtained from the isovector electromagnetic amplitudes through
a VMD analysis. The first column shows the results
employing the helicity amplitudes from Arndt {\it et al} \cite{arndt}. 
In the second column the helicity amplitudes from Feuster
{\it et al} \cite{feupho} are used instead
and the third column shows the results obtained from the  PDG estimates \cite{pdg}. 
For comparison we show also the values for $f_{RN\rho}$ resulting
from direct fits to the partial decay width $\Gamma_{RN\rho}$.
The values for $\Gamma_{RN\rho}$ are
taken from Manley {\it et al} \cite{man2}
(4th column) and PDG (6th column), the $\rho$ decay is parameterized according 
to Eq. \ref{rhowidth1}. To point out further inherent uncertainties in the analysis, we
show in the 5th column the results from a fit to Manley's values for
the partial width, using the parameterization from Eq. \ref{rhowidth2} of the $\rho$
spectral function. }
\label{vmdrho} 
\end{sidewaystable}

\newpage


\begin{table}
\begin{tabular}{|l||c|c|c|c|} \hline
&$\kappa$ & $f_{RN\omega}$(Arndt) & $f_{RN\omega}$(Feuster) & $f_{RN\omega}$(PDG) \\ \hline
$S_{11}(1535)$ &  4       & 1.27$\pm$1.58  & 1.36 & 0.76$\pm$1.23  \\ \hline
$S_{11}(1650)$ &  4       & 1.59$\pm$0.29  & 0.56 & 1.12$\pm$1.09  \\ \hline
$D_{13}(1520)$ &  $8/3$   & 2.87$\pm$0.76  & 2.28 & 3.42$\pm$0.87  \\ \hline
$D_{13}(1700)$ &  $8/3$   & $---$        & 1.88 & 0.65$\pm$2.76  \\ \hline
$D_{13}(2080)$ &  $8/3$   & $---$        &  $---$ & 1.13$\pm$1.46  \\ \hline
$P_{11}(1440)$ &  4       & 0.61$\pm$0.68  & 1.26 & 0.85$\pm$0.48   \\ \hline
$P_{11}(1710)$ &  4       & 0.14$\pm$0.85  & 0 & 0.20$\pm$1.02  \\ \hline
$P_{13}(1720)$ &  $8/3$   & 0.29$\pm$1.30  & 2.18 & 1.79$\pm$3.18  \\ \hline 
$F_{15}(1680)$ &  $4/5$   & 6.89$\pm$1.38  & $----$ & 6.52$\pm$1.49  \\ \hline 
\end{tabular}
\caption{The coupling constants $f_{RN\omega}$ as extracted from the helicity
amplitudes from the Arndt {\it et al} \cite{arndt}(2nd column), 
Feuster {\it et al} \cite{feupho}
(3rd column) and the PDG group \cite{pdg} (4th column).
In the first column we give the values for $\kappa$ (see Eq. \ref{resg}).} 
\label{vmdom}
\end{table}

\newpage

\end{document}